# The crucial role of HST during the NASA Juno mission: a "Juno initiative"

*Paper submitted to the Space Telescope Science Institute in response to the call for HST White Papers for Hubble's 2020 Vision*

March 4, 2015

Grodent, D.[1], B. Bonfond[1], J.-C. Gérard[1], G. R. Gladstone[2], J. D. Nichols[3], J. T. Clarke[4],

F. Bagenal[5], A. Adriani[6]

1 Laboratoire de Physique Atmosphérique et Planétaire, Université de Liège, Liège, Belgium
2 Southwest Research Institute, San Antonio, TX, USA
3 Department of Physics and Astronomy, University of Leicester, Leicester, UK
4 Center for Space Physics, Boston University, Boston, MA, USA.
5 Laboratory for Atmospheric and Space Physics, University of Colorado, Boulder, CO, USA
6 INAF Istituto di Astrofisica e Planetologia Spaziali, Rome, Italy

Contact: Denis Grodent, d.grodent@ulg.ac.be

## Abstract

In 2016, the NASA Juno spacecraft will initiate its one-year mission around Jupiter and become the first probe to explore the polar regions of Jupiter. The HST UV instruments (STIS and ACS) can greatly contribute to the success of the Juno mission by providing key complementary views of Jupiter's UV aurora from Earth orbit. Juno carries an ultraviolet Spectrograph (UVS) and an infrared spectral mapper (JIRAM) that will obtain high-resolution spectral images providing the auroral counterpart to Juno's in situ particles and fields measurements with the plasma JADE and JEDI particle detectors. The Juno mission will be the first opportunity to measure simultaneously the energetic particles at high latitude and the auroral emissions they produce. Following programmatic and technical limitations, the amount of UVS data transmitted to Earth will be severely restricted. Therefore, it is of extreme importance that HST captures as much additional information as possible on Jupiter's UV aurora during the one-year life of the Juno mission. This white paper is a plea for a "Juno initiative" that will ensure that a sufficient number of orbits is allocated to this unique solar system mission.

# 1. Introduction

Magnetospheres are natural laboratories for studying plasmas, the state of matter which represents the vast majority of the visible universe. Jupiter's magnetosphere is the prototype of a 'giant planet', which has become an increasingly important type of planet as more are continually being discovered in the universe. Moreover, Jupiter's magnetosphere acts as a local analogue for more distant astrophysical bodies, such as ultra-cool dwarfs, magnetic white dwarfs and pulsars, and fully exploiting any rare opportunity to maximize our understanding of this object is critical, and will impact across wider astronomical studies.

For almost 25 years, successive ultraviolet (UV) instruments of the Hubble Space Telescope (HST) have been the only instruments able to explore Jupiter's powerful auroral emissions at high spatial and spectral resolution. It is remarkable that after such a long time, new HST observations are still uncovering crucial facts about this multifaceted phenomenon. The complex and highly dynamical auroral morphology provides a global view of the atmospheric response to numerous large-scale and small-scale processes taking place throughout Jupiter's enormous magnetosphere. The understanding of Jupiter's auroral emissions combines several research domains in magnetospheric, atmospheric and satellites sciences. Hence, it fosters the interest of a large community of international scientists, many of whom are now involved in the upcoming NASA Juno mission.

In 2016, Juno will initiate its mission around Jupiter and become the first probe to explore its polar regions. It will answer numerous remaining questions, especially those raised by the stunning HST views of Jupiter's UV aurora. The primary goal of the Juno mission is to explore the origin and evolution of the planet Jupiter [Bolton et al., 2010]. Since Juno's precessing orbit will carry it through more and more of Jupiter's deadly radiation belts, the nominal mission lifetime is only one year. During each of the 33 science orbits Juno will approach and recede from Jupiter through high-latitude magnetic field lines, close to the planet, where significant particle acceleration is expected to take place as a result of its strong magnetosphere-ionosphere coupling. The HST UV instruments (STIS and ACS) can greatly contribute to the success of the Juno mission by providing key complementary views of Jupiter's UV aurora from Earth orbit. The enormous value of simultaneous HST auroral observations with observations from planetary spacecraft has been demonstrated on multiple occasions [e.g. Mauk et al., 2002; Nichols et al., 2007; Radioti et al., 2009, 2011]. The Juno science package includes a UV spectrograph (UVS) that will obtain high quality images and spectra of Jupiter's aurora. However, UVS observations are highly concentrated during each orbit in a few hours near Juno's closest approach to Jupiter. Since one of the main objectives of the Juno mission is to establish the actual connection between the auroral emissions and the particles that generate them [Bagenal et al., 2014], it is of extreme importance that HST captures as much additional information as possible on Jupiter's UV aurora during the one-year life of the Juno mission.

# 2. Juno's Orbits and Relevant Instruments

Juno's orbit insertion at Jupiter will take place in July 2016. After two 53.5-day capture orbits, the spacecraft will perform a series of 33 fourteen-day science polar orbits offering unprecedented views of the auroral regions of Jupiter. The nominal Juno mission will thus overlap HST's Cycles 24 and 25. Juno's highly eccentric science orbits have a perijove close to 5000 km above cloud deck and start out approximately in the Dawn-Dusk plane. The typical operations carried out during the science orbits are illustrated in Figure 1.

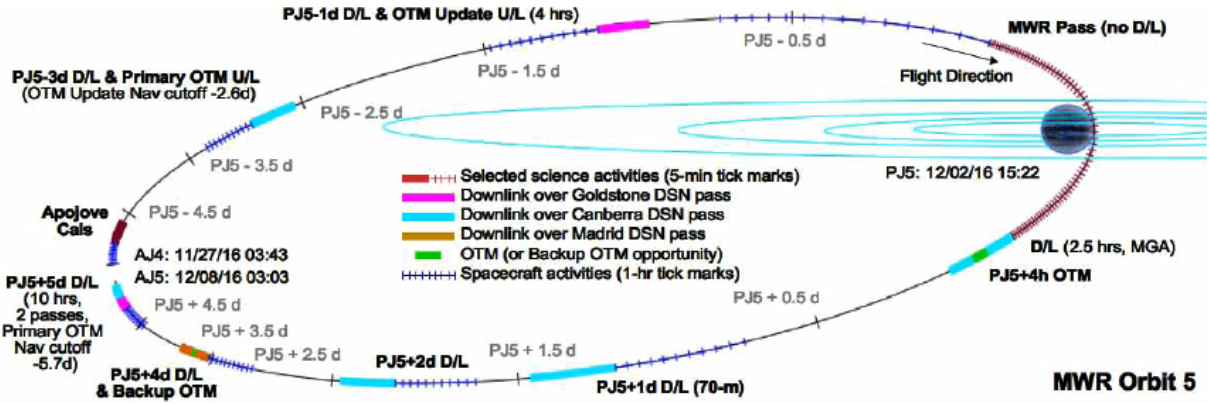

**Figure 1:** One of Juno's 33 science orbits (Dec. 2016). Most of the UVS and JIRAM data will be collected near perijove (red crosses). This 6-hour segment only represents less than 2% of the total duration of one Juno orbit.

Juno carries an UltraViolet Spectrograph (UVS) [Gladstone et al., 2014] and an infrared Imaging Spectrograph (JIRAM) [Adriani et al., 2014] that will characterize the auroral emissions of Jupiter over all science orbits. UVS will obtain high-resolution spectral images that will provide the $H_2$ and H UV auroral counterpart to Juno's in situ particles and fields measurements in the larger polar magnetosphere with the plasma JADE (Jovian Auroral Distributions Experiment) and JEDI (Jupiter Energetic Particle Detector Instrument) detectors. At the same time, the MAG magnetometer and Waves instrument [Bagenal et al., 2014] will accurately constrain magnetic field models and measure signatures of currents and waves flowing along and across the magnetic field lines. The UVS instrument is somewhat similar to the HST STIS G140L spectral slit and is sensitive to EUV-FUV. UVS will take advantage of Juno's stabilizing spin to scan the auroral regions in the direction perpendicular to its slit, while its steerable scan mirror will allow off-spin-plane targeting of specific regions of interest in Jupiter's aurora. The Infrared Auroral Mapper (JIRAM) is equipped with a single telescope and accommodates a near-IR camera and spectrometer. It will observe the $H_3^+$ auroral emissions near 3.4 μm, which overlap a deep methane absorption band. The structure and brightness of this thermal emission depends on both the density of $H_3^+$ ions and the local upper atmospheric temperature. Figure 2 shows a spectral image obtained by reconstructing an HST STIS time-tag sequence [Gérard et al., 2014] during which the G140L spectral slit was scanned across the northern auroral region. This pseudo-image is analogous to the ones that will be reconstructed from the UVS data.

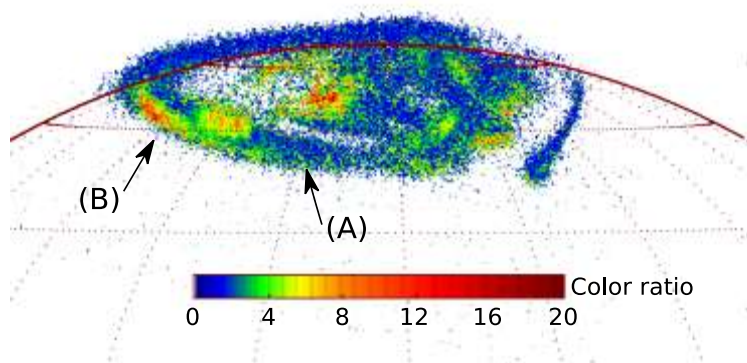

**Figure 2:** Spectral image of Jupiter's northern ultraviolet region reconstructed from an HST STIS time-tag sequence during which the G140L spectral slit was scanned across the aurora (adapted from Gérard et al., 2014). Spectral images obtained with UVS will be similar. The color code refers to the FUV color ratio and is a proxy for the penetration depth of the impinging particles giving rise to the aurora.

Although the nominal Juno mission will take place during the 33 science orbits, much science can already be achieved during Juno's orbital insertion phase. This early period (HST cycle 23) will be the one and only opportunity to obtain a full set of relevant upstream data simultaneously with HST observations, which will determine how Jupiter's magnetosphere is driven by the solar wind.

## 3. Anticipated Science Objectives

Figure 1 highlights the operations during a typical Juno orbit, including the intensive data collection near perijove. The main interest of this segment is that it corresponds to the portion of the orbit during which Juno is flying just above the north and south poles. This viewing geometry is not achievable from Earth and it will provide the first views of the nightside sector of the aurora in both hemispheres. However, this 6-hour segment only represents less than 2% of the total duration of one orbit. High instrumental noise levels due to the high-energy electron radiation may further reduce the amount of usable data. This is where HST is coming into play. In addition to simultaneous and conjugate-region observations near the Juno perijoves, it will be important to observe Jupiter's aurora with HST at other times (when Juno data are sparse), to provide a baseline for the highly variable aurora.

1) When UVS and JIRAM are not observing:
HST STIS and ACS can provide the main characteristics of the auroral emissions that will serve as a global context for the UVS and JIRAM snapshots acquired once every 14 days. Past HST observations have demonstrated that processes taking place in Jupiter's aurora and magnetosphere operate with a wide variety of timescales. As an example, the global internal magnetospheric dynamics of Jupiter is characterized by a typical timescale of several (~6) planetary rotations (e.g. Kronberg et al., 2005; Kimura et al., in press). HST observations carried out before and after the Juno perijove segment are thus necessary to put the Juno observations into the proper magnetospheric perspective. Moreover, the JADE, JEDI, Waves, and MAG instruments will be measuring energetic particles and associated currents along magnetic field lines even outside the perijove segment. Concurrent HST observations would be extremely invaluable to interpret them in terms of connection with the aurora and its spatial extent.

2) When UVS and JIRAM sampling rates are not adequate:
At the other side of the timescale range, auroral and magnetospheric processes are known to vary within 10s of seconds. Juno's spin limits the cadence of UVS and JIRAM observations to every rotation (30s) and telemetry restrictions will prevent long uninterrupted observing sequences. Fortunately, HST has proven that is able to provide 45-minute long and temporally resolved sequences for considerably improved interpretations of the in-situ measurements with JADE, JEDI, MAG and Waves.

3) When UVS and JIRAM are observing elsewhere:
When Juno is very close to Jupiter, it will be impossible for either instrument to monitor the entire auroral oval, or the conjugate auroral region. For the first time, we will be able to observe the whole polar region simultaneously in both hemispheres (e.g. HST looks at the northern aurora and Juno at the southern one). Such observations would be essential to characterize local magnetic field effects and inter-hemispheric electric currents. Exact simultaneity has never been achieved before because STIS and ACS FOVs are too small to observe both hemispheres and because the tilt of the magnetic field prevents simultaneous good views of both hemispheres at the same time.

## 4. Time Allocation Issue

Auroras are, unlike many astronomical and indeed solar system targets, time-domain phenomena. Our understanding of auroral emissions, and thus the dynamics of magnetospheres in general, ultimately derives from observing changes in their properties over all time scales, including medium-long intervals, the latter of which are automatically penalized by the panel review process, which, as experience shows, favors smaller programs. This White Paper raises the issue that a panel review process, which places an extremely heavy emphasis on minimizing orbit numbers, disfavors important science that by necessity requires larger numbers of orbits.

The extraordinary capabilities of HST have the potential to considerably enhance the output of planetary missions and the Juno mission in particular. Such missions, especially those dedicated to the outer solar system, are extremely rare and time critical. When such an opportunity arises, it is important that an appropriately large number of HST orbits are put aside to support these missions without jeopardizing other solar system proposals for this particular cycle.

**We recommend that a category of HST time be allocated specifically for "NASA Juno Mission Support" similar to the "UV initiative", in other words a "Juno initiative".**